\documentclass[conference,10pt]{IEEEtran}
\usepackage{tabularx}
\usepackage{algorithm, algpseudocodex}
\usepackage{bytefield}
\usepackage{amsmath}
\usepackage{amsthm}
\usepackage{graphicx}
\usepackage{subcaption}
\usepackage{dblfloatfix} % for floating figure*
\graphicspath{{figures/}}

\usepackage{hyperref}

\title{zk-PoT: Zero-Knowledge Proof of Traffic \\
for Privacy Enabled Cooperative Perception}
\author{
  \IEEEauthorblockN{
    Ye Tao \IEEEauthorrefmark{1},
    Yuze Jiang \IEEEauthorrefmark{1},
    Pengfei Lin \IEEEauthorrefmark{1},
    Manabu Tsukada \IEEEauthorrefmark{1},
    Hiroshi Esaki \IEEEauthorrefmark{1}
  }
  \IEEEauthorblockA{
    \IEEEauthorrefmark{1} Graduate School of Information Science and Technology, The University of Tokyo \\
    \IEEEauthorrefmark{1} \{tydus, yuzu, pengfei, tsukada\}@hongo.wide.ad.jp, hiroshi@wide.ad.jp
  }
}
\begin{document}

\maketitle

\begin{abstract}
    Cooperative perception is an essential and widely discussed application of connected automated vehicles. 
    However, the authenticity of perception data is not ensured, because the vehicles cannot independently verify the event they did not see.
    Many methods, including trust-based (i.e., statistical) approaches and plausibility-based methods, have been proposed to determine data authenticity.
    However, these methods cannot verify data without a priori knowledge.
    In this study, a novel approach of constructing a self-proving data from the number plate of target vehicles was proposed.
    By regarding the pseudonym and number plate as a shared secret and letting multiple vehicles prove they know it independently,
    the data authenticity problem can be transformed to a cryptography problem that can be solved without trust or plausibility evaluations.
    Our work can be adapted to the existing works including ETSI/ISO ITS standards while maintaining backward compatibility.
    Analyses of common attacks and attacks specific to the proposed method reveal that most attacks can be prevented, whereas preventing some other attacks, such as collusion attacks, can be mitigated.
    Experiments based on realistic data set show that the rate of successful verification can achieve 70\% to 80\% at rush hours.
\end{abstract}


\section{Introduction}
\label{sec:intro}
Road transportation has been one of the most essential services for human mobility since ancient times.
But since then, the primary form of transportation has remained almost unchanged: depending on the environment, the driver of the vehicle
determines how to drive.
Current vehicles are equipped with sensors including LiDAR, mmwave, stereo camera. According to sensor data, driving assistant capabilities and autonomous driving can be realized.
However, sensors cannot overcome locality limitations. Because such sensors are mounted on the vehicles, in theory, the sensors share a similar view as the driver's vision. Therefore, these sensors cannot see too far or beyond obstacles as well.

Connected autonomous vehicles can perform numerous tasks with improved computational power on embedded systems and advanced technologies such as deep neural networks. Furthermore, various low-latency vehicle-to-everything (V2X) communication methods, including 5G and vehicular ad-hoc network (VANET) \cite{Etsi2014-fu}, render real-time vehicle cooperation possible. With such technological advances, cooperative perception has evolved rapidly \cite{Gunther2015-js}.

Cooperative perception is one of the most revolutionary inventions in autonomous driving, not only because other vehicles' locations can be acquired from their broadcasting messages but also other vehicles' sensors can be used to enlarge our view and reduce blind spots. Cooperative perception is standardized in ISO \cite{International_Standard_Organization2020-wf} and ETSI \cite{Etsi2010-hk}.

\begin{figure}[ht]  \includegraphics[width=\linewidth]{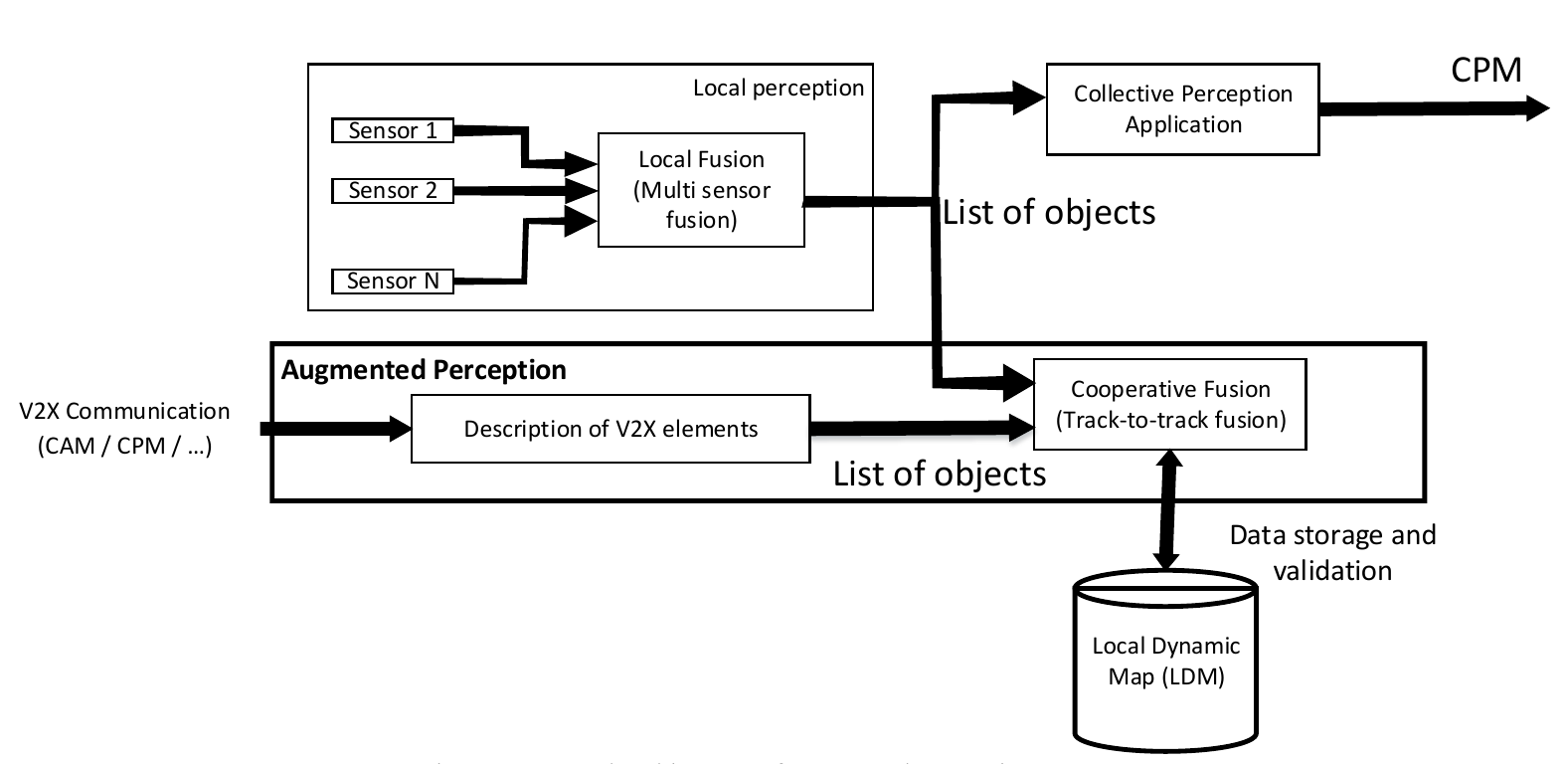}
  \caption{General Structure of Collective Perception Service (CPS) \cite{Merdrignac2018-sl}} \label{fig:cps}
\end{figure}

Collective perception service (CPS) is defined as a critical safety-related application protocol in the ITS station architecture standardized in ETSI, as displayed in figure~\ref{fig:cps}. Collective perception message (CPM) is the format used to share observed vehicles with others.
CPM contains the position, and velocity, with other information about the observed vehicles.

Connected vehicles that receive these messages would make decisions based on them. Therefore the authenticity of the CPM is crucial for road safety \cite{Bian2018-es}.
Because the CPM is generated by vehicles and not by a centralized and trusted authority, the authenticity of the data is a vital concern \cite{Hasrouny2017-xb,Golle2004-px}.
In contrast to the authority, vehicles could be motivated to lie to the other vehicles to achieve their profits \cite{Leinmuller2008-uw}.

Many solutions have been proposed to identify and nullify liars.
For example, public key infrastructure (PKI) has been standardized in VANET \cite{Etsi2021-bq}, which requires every message in  VANET should be signed with the secret key provided by the law enforcement agencies (LEAs).
Thus, every node in the network is forced to obtain the keys from LEA's before sending any message.

However, vehicles can still send malicious messages intentionally and sign with their keys.
Because of the highly decentralized and mobile nature of VANET, existing security solutions are not feasible. 
Many trust management methods have been developed to overcome this problem \cite{Minhas2011-bw,Raya2008-ja,Yang2019-oj,Suo2019-no}.
These methods depend on estimations of trustworthiness.
Trust management schemes in VANET can be categorized into data-centric and
entity-centric methods. However, both of these methods exhibit deficiencies:
in data-centric methods, a large amount of data is needed, which leads to packet loss and increased latency;
also in many cases, missing or hard-to-obtain ground truth makes the approaches only able to rely on ``feasibility'' evaluations;
in entity-centric methods, data correctness remains an issue due to the existence of attackers and limitations of sensors\cite{Soleymani2015-ik}.

% Contribution
In this study, we proposed the zero-knowledge Proof of Traffic (zk-PoT), a proof-based traffic verification scheme.
By letting vehicles zero-knowledge-prove the observation of the target vehicles independently, the ego vehicle can be immediately convinced of the existence of the target vehicle without the requirement of trust evaluation or feasibility evaluation. 

This method can preserve the privacy of the target vehicles; furthermore, the overall security, efficiency, and data latency can be improved.
This method can be integrated into the existing cooperative perception standards, including ISO and ETSI, requiring little changes to the architecture, and maintains the backward compatibility.
It can also be used in conjunction with existing trust management schemes as a bootstrap.

The rest of the paper is organized as follows:
In section~\ref{sec:related}, we highlight the dilemma of location privacy versus trust management. Next we shed some light on the cryptographic tools necessary to approach our solution: elliptic curve digital signature algorithm (ECDSA) and zero-knowledge proofs.
In section~\ref{sec:problem}, the problem is defined, reasonable assumptions were made, and the approach to the problem was discussed.
In section~\ref{sec:proposal}, the problem was transformed into a cryptographic problem, which is then solved using zero-knowledge proofs. Finally, this solution can be applied by extending the existing CPS protocol.
In section~\ref{sec:evaluation}, the resilience of the proposed solution against some attacks was analyzed, and a quantitative analysis based on our proof-of-concept system was conducted.
Finally, in section~\ref{sec:conclusion}, the contributions of the proposed method were summarized and future work was discussed.

\section{Related Work}
\label{sec:related}
\subsection{Location Privacy and Trust Management Dilemma}
Compared with conventional vehicles, in which limited
to no information is reported about their position and velocity, connected autonomous vehicles (CAVs)
are designed to share their information with other CAVs on the road.
Most trust management methods require tracking vehicles' historical behaviors. Thus, information is exposed to the trackers.
Therefore, guaranteeing accurate trust evaluation and simultaneously preserving location privacy is difficult.
For example, a pseudonymous authentication scheme is can be used to protect vehicles' location privacy \cite{Mansour2018-tz}; however, trust management becomes challenging because tracking a specific vehicle becomes difficult.

\subsection{Digital Signature and ECDSA}
A digital signature is a mathematical scheme for verifying the authenticity of digital messages or documents. Asymmetric cryptography is used to allow an entity to verify the authenticity of a message.
Digital Signature Algorithm (DSA) is a standard of digital signature, which is based on the difficulty of the discrete logarithm problem.
ECDSA \cite{Johnson2001-sx} is a variant of DSA which uses elliptic-curve cryptography.
Compared with other digital signature algorithms, ECDSA exhibits distinctive properties.
Specifically, among them, two properties are closely related to our problem.

\newtheorem{prop}{Property}
\begin{prop}
\label{prop:1} Private key is plain integer
\end{prop}

Compared with RSA, which generates public-private key pair from a pair of large primes, the private key of DSA (including ECDSA) is just an integer. This property enables generating key pairs from arbitrary data.

\begin{prop}
\label{prop:2} Public key can be recovered from signature
\end{prop}

Given a signed message and the corresponding signature, the public key used in the signing process can be derived (i.e., recovered). This property enables omission of the transmission of the public key, which conserves the bandwidth of V2V communication.

\subsection{Zero-Knowledge Proofs}
The zero-knowledge proof system was initially proposed in \cite{Goldwasser1989-wy} in 1989.
In such a system, one party, prover, can convince the other party, called the verifier, that a specified statement is true without revealing any other information to the verifier besides that the statement is true.

A zero-knowledge proof system should satisfy the following three requirements:

\begin{enumerate}
    \item \textit{Completeness}: If the statement is true, the verifier can be convinced by the prover.
    \item \textit{Soundness}: If the statement is false, the verifier cannot be convinced by the prover, even if the prover cheats.
    \item \textit{Zero-knowledge}: The verifier cannot gain any knowledge apart from the statement being true.
\end{enumerate}

Zero-knowledge proof systems are widely used by various cryptocurrency protocols, such as Zerocoin \cite{Miers2013-za} and Zerocash \cite{Ben_Sasson2014-ka}.
However, this technique is new to other fields of study, including ITS.

\section{Problem Statement, Assumptions and Approaches}
\label{sec:problem}
\subsection{Problem Statement}
Although the CPM can facilitate cooperative perception ability, the riskiness of such messages should be considered.
We cannot blindly trust CPM, otherwise, vehicles could make non-optimal or even dangerous decisions, which jeopardizes road safety.
Previous studies have typically relied on trust estimation and management \cite{Minhas2011-bw}.
However, these methods are based on previous statistics. Therefore, recognizing one-shot lies is almost impossible.
By contrast, we designed a cryptographic scheme that enables vehicles to prove their observations of the traffic.
In this method, genuine observations can be endorsed.

Among the various events that occur on the road, ``existence'' is the most fundamental and accurate event, which does not change because of the high speed of vehicles, as opposed to the location of a vehicle, which will quickly become inaccurate after only a short time. 
Therefore, in this work, we concentrate on proving the existence of observed vehicles.

% This figure belongs to section proposal.
\begin{figure*}[t]
    \centering
    \includegraphics[width=.75\linewidth]{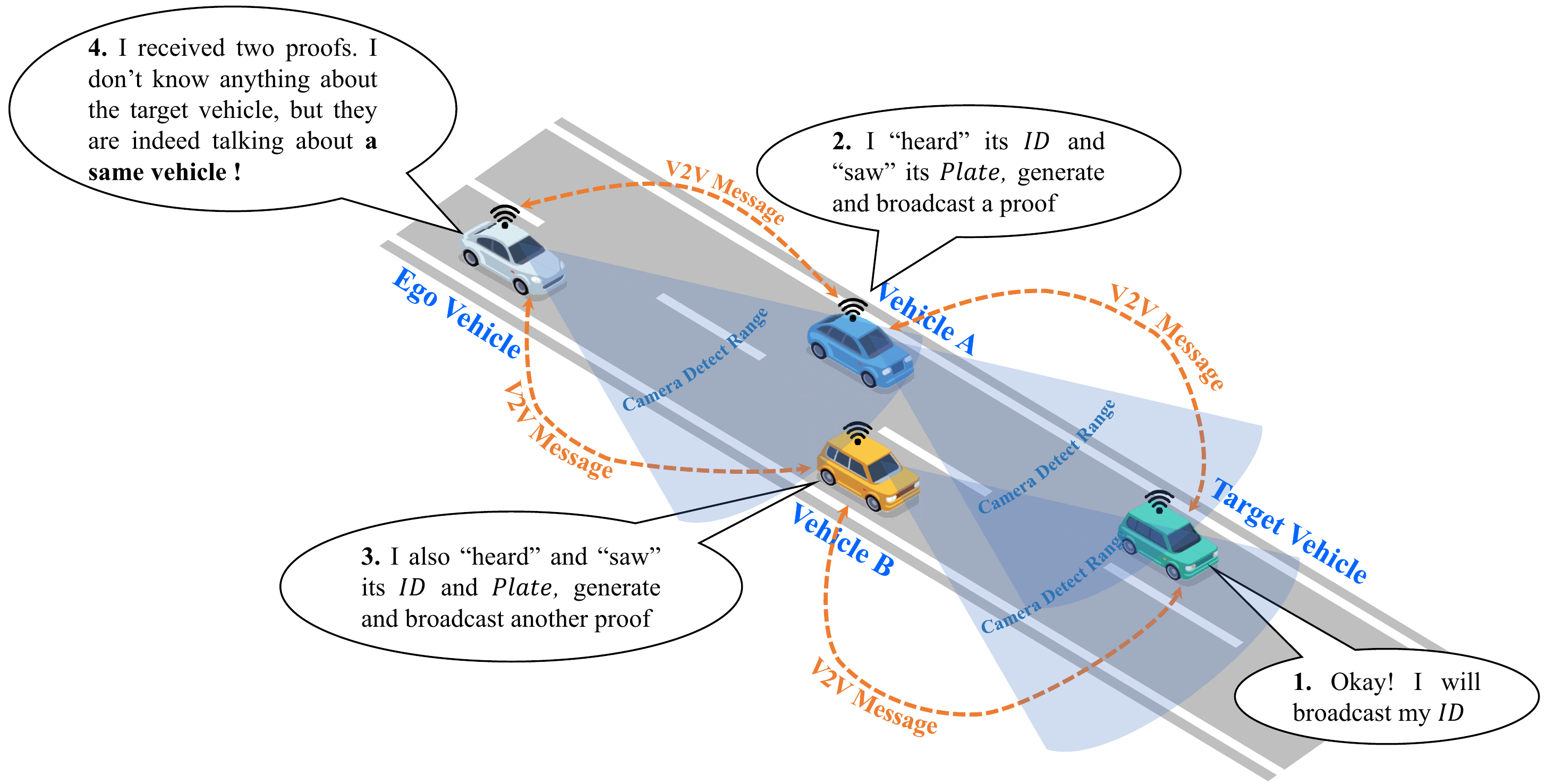}
    \caption{Procedure of a zero-knowledge proof}
    \label{fig:scenario}
\end{figure*}

\subsection{Assumptions}
Every vehicle on the road is assumed to be joining VANET, then all vehicles can perceive and share the perception with other vehicles.
Therefore, we do not distinguish between connected vehicles and non-connected vehicles later in this paper. 
PKI is supported for all infrastructures and all vehicles.
LEAs distribute key pairs to all vehicles.
Vehicles sign each message they send and use the public key to verify the authenticity of the message they received.
A pseudonym system is assumed to be used with perfect untraceability, that is, after a vehicle changes its pseudonym, the new pseudonym cannot be linked with the old pseudonym.

According to the assumptions, we define ``A heard B'' as A has received a message containing B's pseudonym;
we define ``A saw B'' as A has identified the number plate of B.
Because the V2V communication range is designed further than the sensors' vision range, if A sees B, A could surely hear B.

\subsection{Approaches}
A na\"ive ``proof system'' could let vehicles broadcast their observed vehicles' number plates.
However, this method not only destroys the location privacy of the observed vehicles but is also exposed to replay attacks and data forging.

A zero-knowledge proof (ZKP)-based proof system was introduced to solve both privacy and forging problems.
Specifically, we let the vehicles prove they have seen another vehicle.
Using the assumption of ``seeing'' and ``hearing,'' the vehicle that can link the number plate and pseudonym should have seen the target vehicle, which is strong evidence of the existence of the target vehicle.
Moreover, both pseudonyms and number plates cannot be disclosed from the proofs as long as our proof system satisfies the zero-knowledge property.

% What is the problem of zkp
However, zero-knowledge proofs exhibit a limitation: the verifier cannot falsify the proof, which means to ``prove'' that the proof is not true.
This limitation is directly derived from the zero-knowledge property: zero or limited information (i.e., entropy) about the statement can be deduced from the proof, so too many true or false possibilities exist.
Thus, proof of a vehicle's existence cannot be directly falsified by any party, even the parties that know the ground truth.

To solve this problem, we proposed a novel approach by letting different vehicles make their own proofs to the same target vehicle.
Any third party obtaining more than one proof can easily verify if two proofs exist for the same vehicle without knowing any information about the particular vehicle.

\section{Proposed Method}
\label{sec:proposal}
To let vehicles prove they see the same vehicle, the problem was converted to a cryptographic model named \textbf{zero-knowledge proof of shared secret (zk-PoSS)}. Next, we provided a solution to this model. Finally, the proposed model was applied to the existing ETSI standard by extending the packet structure and station behaviors. We chose to use ETSI standards as an example, but this study can also be applied on other cooperative perception systems, such as ISO standards.
\begin{figure*}[b]
    \centering
    \includegraphics[width=\linewidth]{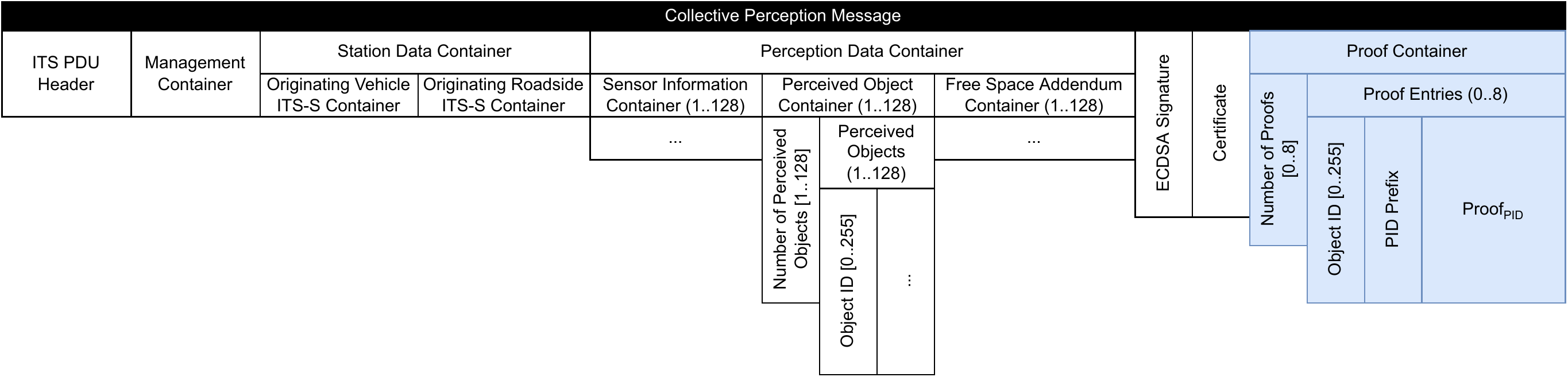}
    \caption{Structure of the extended collective perception message (CPM) with proofs}
    \label{fig:cpm-extended}
\end{figure*}

\subsection{Problem Conversion}

As displayed in figure~\ref{fig:scenario}, two vehicles, A and B, observe a vehicle named Target, and they acquired its pseudonym, $ID$, and number plate, $Plate$.
According to the observed $ID$ and $Plate$, two parties, A and B, essentially obtain a \textit{shared secret}, which can be derived from the target vehicle's $ID$ and $Plate$.
Thus, if ``evidence'' to the fact that they indeed know the \textit{shared secret} could be published, everyone could be convinced without direct observation of the Target.
Additionally, to protect the privacy of the Target and against simply replaying, the ``evidence'' should not disclose the plaintext of $ID$ and $Plate$.

Thus, our problem was transformed to a cryptography model called \textbf{zero-knowledge proof of shared secrets (zk-PoSS)}, which can be formalized as follows:

Given a shared secret $SS$ and two cryptography salts $salt_a$ and $salt_b$, two proofs can be made by the one-way \cite{Robshaw2011-ww} proof function $F_{proof}$.
Next, another function $F_{verify}$ can be feed with $(P_a, salt_a, P_b, salt_b)$ and return if the pairs are about the same $SS$.

\begin{align*}
    &P_a = F_{proof}(SS, salt_a) \\
    &P_b = F_{proof}(SS, salt_b) \\
    &P_a \neq P_b \\
    &F_{verify}(P_a, salt_a, P_b, salt_b) = True
\end{align*}

\subsection{Zero-Knowledge Proof of Shared Secret}
Using ECDSA, \textbf{zk-PoSS} can be solved.
Because we have a shared secret with sufficient entropy, (hash of) the secret can be used as a private key (ECDSA property~\ref{prop:1}).
Then, by definition, a valid signature becomes evidence (i.e., proof) that the prover holds the private key.
In the proposed scheme, holding the private key means the prover indeed knows the shared secret. 
Exploiting such a property of digital signature, two parties make their own proofs of the shared secret, using different random salts.
Next, a third party who has both proofs can recover the public keys from the signatures (ECDSA property~\ref{prop:2}), and check if the two public keys are the same.

When public keys are the same, private keys are the same, and the private key is essentially a hash of the shared secret. 
Thus, the only possibility is that they indeed have the same shared secret.

\hfill

The proving system is detailed as follows:

{\setlength{\parindent}{0pt}
\subsubsection{Setup}
\begin{itemize}
    \item Every party agrees with the same set of ECDSA Parameters: $CURVE, G, n$.
    \item Every party agrees with the same cryptography hash algorithm $H$, which produces fixed-length output with the same binary length of $n$.
\end{itemize}

\subsubsection{Make a proof}
\begin{itemize}
    \item Calculate hash of shared secret with a cryptographic hash function $h = H(SS)$.
    \item Represent the hash as an unsigned integer $sk$. If $sk >= n$, retry first step and instead calculating hash $h == H(SS||SS)$ until $sk < n$.
    \item Calculate $PK = G \times sk$ on curve $CURVE$. $sk$ and $PK$ can be seen as a pair of secret key and public key of the ECDSA algorithm, respectively.
    \item Sign a random message $M$ with $sk$: $Sig(M)$.
    \item Publish $M||Sig(M)$. $||$ denotes binary concatenation.
\end{itemize}

\subsubsection{Pair and verify proofs}
\begin{itemize}
    \item Recover $PK$ from $M$ and $Sig(M)$ and store it in database.
    \item If the database contains another message $M'$ with the same $PK$, then it has very high confidence (with a false positive rate of $\frac{1}{n}$) in both messages and considers these messages highly trustworthy.
\end{itemize}
}

zk-PoSS is a zero-knowledge proof system because it satisfies all three properties:

\begin{enumerate}
    \item \textit{Completeness}: If the $PK$s are same, $sk$ (i.e., $SS$) should be same. Thus, the verifier should be convinced.
    \item \textit{Soundness}: If the $SS$s are not the same, only a hash collision, which is difficult to construct on a cryptographic hash function, can produce the same $sk$. Thus, a cheating prover has negligible chance of convincing the verifier.
    \item \textit{Zero-knowledge}: The verifier cannot deduct any $sk$ from $PK$. This property is guaranteed by any public key cryptography, including ECDSA.
\end{enumerate}

\subsection{Update to the CPS}
% Detail the actual algorithm and consequential protocol update.

Using zk-PoSS, the CPS can be extended to support proof of traffic capability.
We select the standardized \texttt{secp256k1} curve \cite{Brown2010-ar} which is an 256-bit curve and the signature is 65-byte long.
The message $M$ used in zk-PoSS is the prover's pseudonym.
It was chosen under three considerations: it provides enough entropy; it is transmitted in every packet, so we can omit it to save bandwidth; it helps prevents the na\"ive replay attack, which will be discussed later in section~\ref{sec:evaluation}.

\subsubsection{Changes to the CPM packet structure}
As depicted in figure~\ref{fig:cpm-extended}, the CPM was extended with an additional section named \texttt{Proof Container} in the end, which contains a list of 0 to 8 \texttt{Proof Entries}.
Each \texttt{Proof Entry} includes an \texttt{object ID} that appears in the perceived object container, a 32-bit prefix of the vehicle's ID to facilitate quick filtering, and actual proof of this ID.

\hfill 

\begin{figure}[h]
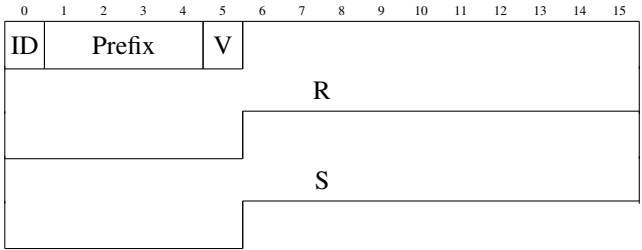

    \centering
    \begin{bytefield}[bitwidth=1.5em]{16}
        \bitheader{0-15} \\
        \bitbox{1}{ID} & \bitbox{4}{Prefix} & \bitbox{1}{V} & \bitbox[tlr]{10}{} \\
        \wordbox[lr]{1}{R} \\
        \bitbox[lrb]{6}{} & \bitbox[tlr]{10}{} \\
        \wordbox[lr]{1}{S} \\
        \bitbox[lrb]{6}{} & \bitbox[lt]{10}{}
        
    \end{bytefield}
    \caption{Packet structure of proof entry (70 bytes)}
    \label{fig:proof-entry-structure}
\end{figure}

The detailed packet structure of the proof entry is depicted in figure~\ref{fig:proof-entry-structure}. \texttt{ID} denotes the \texttt{object ID}, \texttt{Prefix} denotes the 32-bit prefix, and \texttt{V}, \texttt{R}, \texttt{S} are the three variables used in ECDSA signatures respectively \cite{Brown2009-bu}.

Given that the CPM is sent in a high frequency (up to 10 Hz \cite{Etsi2019-mg}) and the proof extended by this work could be lengthy (70 bytes per \texttt{Proof Entry}), to save bandwidth, the proofs should be sent occasionally and omitted in the other CPMs.
The client should store each entry for a time ahead.
The specific parameter is not decided yet but once per 3 s should be sufficient to handle VANET topology changes.

\subsubsection{Changes to the ITS Station}

\begin{figure*}[htb]
     \centering
     \begin{subfigure}[b]{0.49\linewidth}
         \centering
         \includegraphics[width=\textwidth]{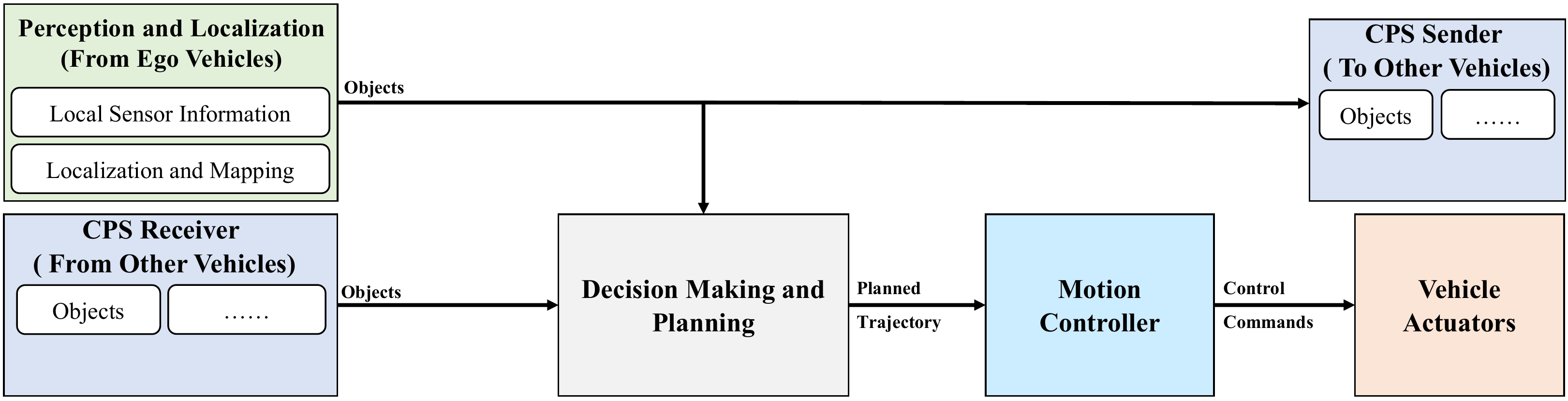}
         \caption{Original CPS}
         \label{fig:cps-original}
     \end{subfigure}
     \hfill
     \begin{subfigure}[b]{0.49\linewidth}
         \centering
         \includegraphics[width=\textwidth]{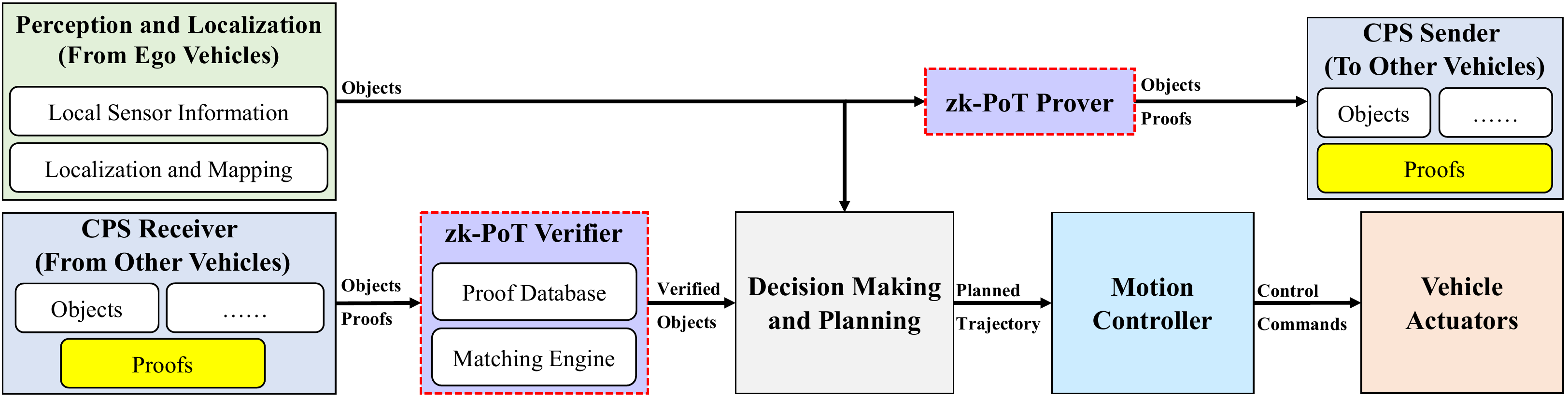}
         \caption{Extended CPS}
         \label{fig:cps-extended}
     \end{subfigure}
     \caption{Block Diagram of Original Versus Extended CPS}
     \label{fig:cps-original-extended}
 \end{figure*}
 
Although different implementations could result in distinct module designs, the conceptual structure of a vehicle with CPS should be similar to that in figure~\ref{fig:cps-original}.
The \texttt{receiver} module receives objects from CPM, feeds it together with \texttt{local} perceived objects into \texttt{planning} module for the next steps.
Furthermore, the objects from the \texttt{local} module should be copied to the \texttt{sender} module for CPM generation and sending.

We extended the CPS by inserting two new modules: \texttt{prover} and \texttt{verifier}.
As depicted in figure~\ref{fig:cps-extended}, the \texttt{prover} module receives the objects from \texttt{local}, generates proof for each possible objects, and subsequently outputs the proofs with the original objects to the \texttt{sender} module to be staged in CPM and sent out.
By contrast, instead of directly taking all received objects into account, the \texttt{verifier} module is inserted after \texttt{receiver} and works as a gatekeeper.
It receives ``raw'' objects and proofs, stores them in its internal database, and continuously matches the proofs.
When a match is found, the corresponding object is considered trustworthy.
All recent data, including the one staged in the internal database, should finally be sent to the \texttt{planning} module.
All consequent data from the same object should also be sent immediately to the \texttt{planning} module.

The changes to the modules already designed and implemented by such design can be minimized. Furthermore, backward compatibility to the standards is maintained.

\section{Analysis and Evaluation}
\label{sec:evaluation}
\subsection{Threat Analysis}
As proposed, with the power of zk-PoT, vehicles can immediately verify the data received from CPS.
Thus, na\"ive attacks and data manipulation (including generating fake vehicles, reusing known vehicles, etc.) can be reliably filtered out.
Furthermore, different complicated attacks could also be blocked or mitigated.
This list is not a comprehensive list of possible attacks, but within the limited space, the following common attacks are discussed.

\subsubsection{Brute-Force Attack and Dictionary Attack}
Malicious parties can brute force the number plate used in the proof to match the public keys used in other vehicles' proofs if only they ``heard'' the vehicle.
Moreover, this brute-force attack can be conducted in a fully offline fashion: no information exchanged with other vehicles is required.
Key derivation functions (KDFs), such as PBKDF2 \cite{Kaliski2017-gr} or scrypt \cite{Percival2016-px}, can be introduced to extend the brute-force time to be longer than the longest lifetime of a pseudonym. Thus, this attack is mitigated.

In addition to brute-force attacks, malicious vehicles can collect the previously known number plates and use them as a dictionary.
Given the locality of vehicles, the success rate can be higher than those of brute-force attacks.
Some (ephemeral) salt mechanics should be considered to address this issue.
% Introduce random salt mechanism.

\subsubsection{Location Privacy Against Tracking}
Location privacy is essential in V2V communications, which is why pseudonyms are invented.
Given that the pseudonym and KDFs are used to generate the proof, a timely proof cannot be created by brute-force attack, as mentioned.
However, attackers could still collect the data on the road and eventually reveal the trajectory of a vehicle heard.
Despite this, after pseudonym ID changes, remote attackers could no longer track its plate even though they know its old $ID$ and $Plate$.
Thus, the proposed mechanism can provide at least the same level of privacy as the original pseudonyms.

\subsubsection{Spam Attack}
Because the partial proofs cannot be falsified, verifiers are exposed to spam attacks.
Specifically, the attackers can easily generate random ``valid-looking'' proofs, which waste the computing power and memory of the victim verifiers and could result in denial-of-service.
This type of attack can be mitigated by limiting the number of unmatched proofs from each prover.
When a vehicle exceeds the quota, a verifier should silently ignore any consequent proofs sent by this vehicle until some proofs are matched by other vehicles.

\subsubsection{Signal Jamming Attack}
In the proposed scheme, a malicious vehicle could not simply replay the message.
However, an attack can still be performed by jamming the signal from the original prover and subsequently replaying its proof on its own behalf.
Our countermeasure to this type of attack by including the prover's own $ID$ in the proof, because $ID$ is not easily forgeable.

\subsubsection{Collusion Attack and Sybil Attack}
% organized (i.e. collusion attack) => 1/N(A, B)
% Make marginal revenue shrink (sharply)
% and limit the number of pseudonyms (i.e. introduce a financial system)
A collusion attack is a type of attack specific to the method that requires cross verification, including the proposed method.
Sybil attack \cite{Douceur2002-re} is an attack in which an attacker creates a number of pseudonymous identities and uses them to pretend to be different users.
These two types of attacks can be used simultaneously to generate proofs of a set of the same fake vehicles, which can, in turn, convince honest verifiers.

These types of attacks are very difficult to be solved generally, but we have some approaches to mitigate them.
For the Sybil attack, we can set up mechanisms to prohibit the simultaneous usage of different pseudonyms belonging to the same vehicle.
The design of pseudonyms is not directly related to the proof system itself and will be discussed in the future.
For the (non-Sybil) collusion attack, the number of vehicles controlled by an attacker is limited (i.e., they need a platooning fleet).
In this case, we can limit the marginal revenue of the data.
Depending on the system to be built based on the proof system, the ``revenue'' could be a trust value and some credits.
For example, we can set the trust value of one successful match by vehicles A and B to be the reciprocal of the number of accepted matches in the past 15 min.
This countermeasure can cause the trust decrease sharply, so further collusion will not be profitable.

\subsection{Proof-of-Concept Based Evaluation}

\begin{figure}[t]
    \centering
    \includegraphics[width=.9\linewidth]{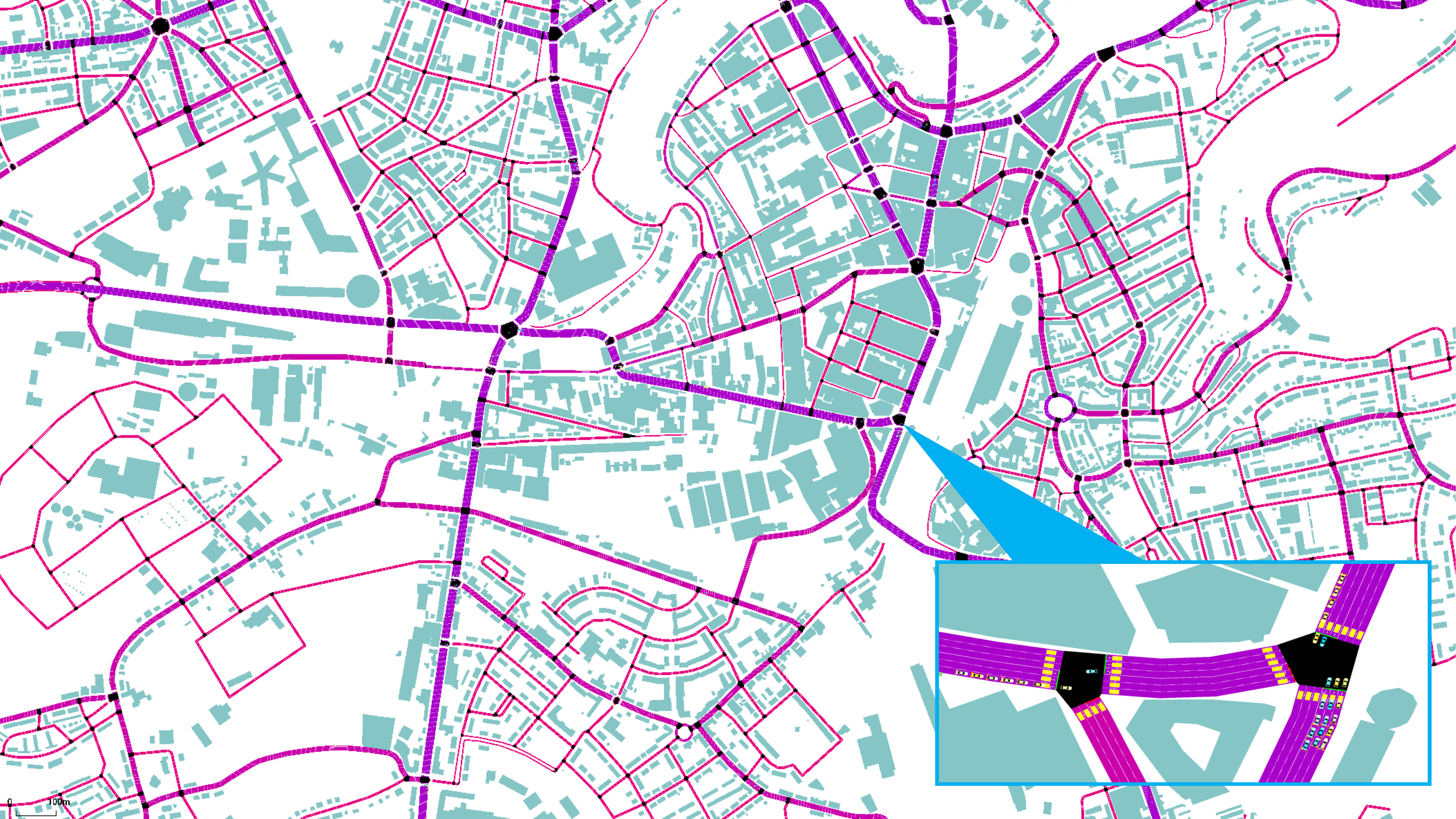}
    \caption{LuST data set in Luxembourg used for evaluation in the SUMO simulator}
    \label{fig:lust}
\end{figure}

%In this section, we have evaluated the proposed zk-PoT method.
We used the 24-hour realistic traffic data generated from Luxembourg \cite{Codeca2018-pu} in the SUMO simulator (Simulation of Urban MObility), which is a highly portable, open source, microscopic and continuous multi-modal traffic simulation package designed to handle large networks \cite{Lopez2018-ps}.
As shown in figure~\ref{fig:lust}, the data set nearly covered the entire city and recorded the traffic flow in different scenes, including intersections and roundabouts.
Furthermore, the LuST data set contains the morning (from 7:00 to 9:00) and evening (from 17:00 to 19:00) peak traffic flows that can verify the effectiveness of the proposed method for large traffic flow.
The data set also provides and evaluates a scenario that meets all the standard requirements, including size, realism, and duration.

A proof-of-concept implementation of zk-PoT was built and used to evaluate the performance in the Luxembourg scenario.
First, we collected floating car data (FCD) from SUMO, which contains the position and angle information of every vehicle at each tick.
Then, we fed this data to our program, which simulates the camera view of vehicles.
The visibility of other vehicles' number plates is tested in the simulation.
For each number plate a prover vehicle observed, a proof to the vehicle which is corresponding to pseudonym and number plate is generated and broadcasted.
Next, other vehicles continuously match and validate the proofs they received.
Finally, we collect the data from the vehicles as our metrics.

\begin{table}[ht]
    \centering
    \caption{Parameter Settings}
    \label{tab:para}
    \begin{tabular}{ccc}
        \hline
        Parameter& Value& Unit\\
        \hline
        Vehicle Length& 4.0& meter\\
        Vehicle Width& 1.8& meter\\
        Perception Distance& 65.0& meter\\
        Camera Sensing Angle& 120.0& degree\\
        Communication Range (each vehicle)& 300& meter\\ 
        Communication Delay (each vehicle)& 1& ms\\
        \hline
    \end{tabular}
\end{table}

For the simulation settings, the following table~\ref{tab:para} illustrates the detailed parameters selection.
Note that our evaluation only concentrates on sedan cars and ignores other vehicle types (e.g., trucks and vans).
Additionally, since the traffic flow during the midnight hours is relatively sparse, we only parse the LuST data from 4:00 to 24:00.
The basic scenario setting is depicted in figure~\ref{fig:scenario} where the target vehicle is occluded by vehicle A and vehicle B from the view of the ego vehicle, and a pair of proofs from A and B will help.
We extend this scenario to large traffic flow to evaluate the macroscopic effectiveness of the zk-PoT based on three significant performance indexes: confirmed vehicle rate, confirmed proof rate, and the number of proofs announced simultaneously for a single vehicle. 

\begin{figure}[ht]
    \centering
    \includegraphics[width=.9\linewidth]{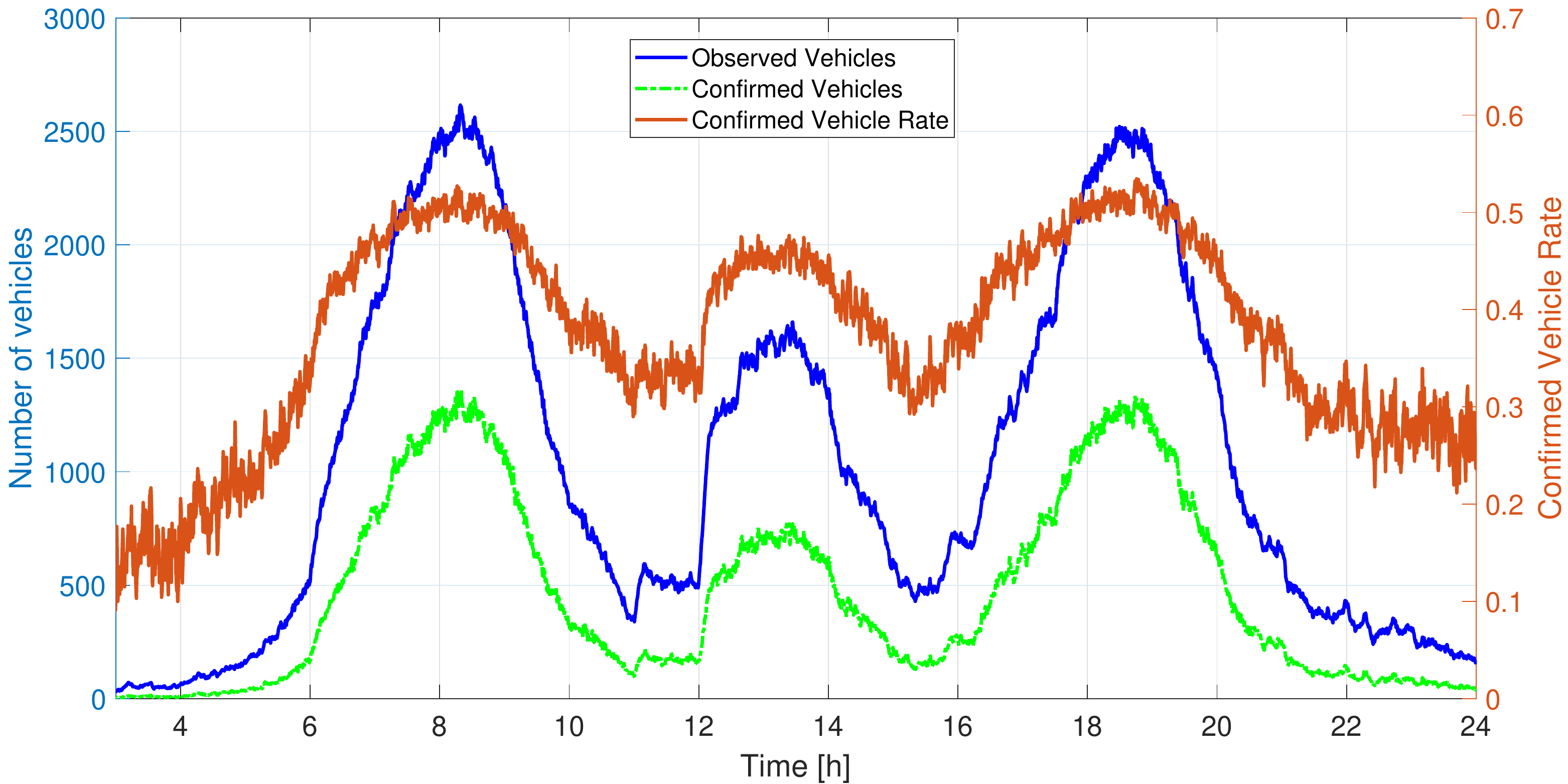}
    \caption{Numbers of observed vehicles and confirmed vehicles, as well as confirmed vehicle rate}
    \label{fig:vehicle_data}
\end{figure}
\begin{figure}[ht]
    \centering
    \includegraphics[width=.9\linewidth]{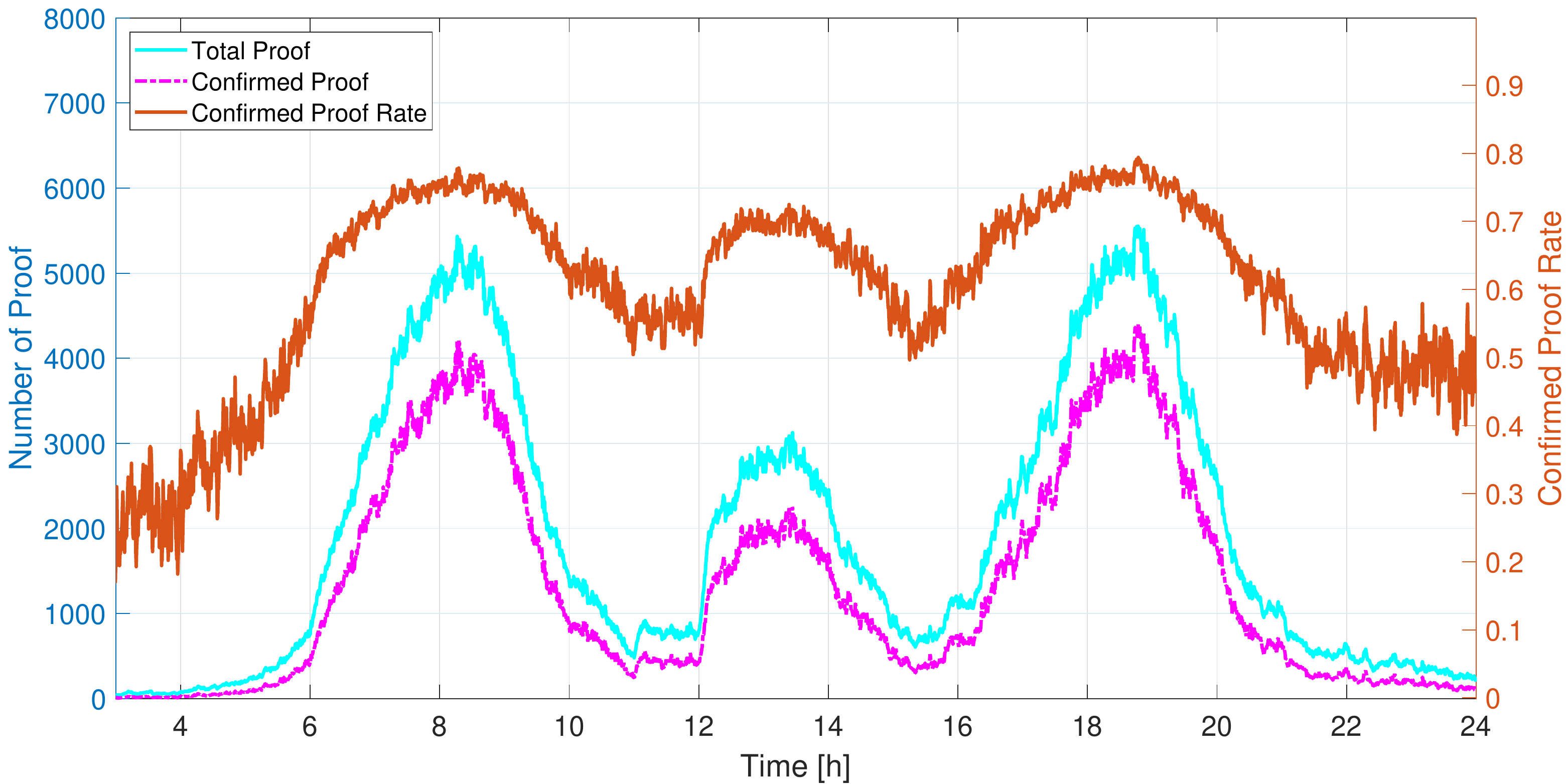}
    \caption{Numbers of total proof and confirmed proof, as well as confirmed proof rate}
    \label{fig:proof_data}
\end{figure}

The numbers of observed vehicles and confirmed vehicles are described in figure~\ref{fig:vehicle_data}. We can observe three peaks of the traffic flow (the solid blue line) at around 8:20, 13:25, and 18:30, respectively. The peak values are 2605, 1660, and 2602, respectively. Besides, the numbers of confirmed vehicles (the green dashed line) at the aforementioned peaks are 1356, 774, and 1321, respectively. Therefore, the corresponding confirmed vehicle rate is denoted as the solid brown line where we can compute the average confirmed vehicle rate (45.9\%) around the peak times.
On the other hand, the generated and confirmed proof numbers are presented in figure~\ref{fig:proof_data}. We can see that the number of proofs grows with traffic volume; the total proofs (the solid cyan line) reach 5433, 3125, and 5513, respectively, at each peak. In the meantime, the numbers of confirmed proof (the magenta dashed line) from each peak are 4204, 2250, and 4404, respectively. Therefore, we can observe high values in the confirmed proof rate from each peak which are 77.85\%, 72.38\%, and 79.39\%, respectively. Besides, the average confirmed proof rate is nearly 70.26\% which explicitly states that the performance of our proposed method maintains a high level, especially in the morning and evening peak times when visual occlusions are likely to happen, which may lead to accidents.

\begin{figure}[ht]
    \centering
    \includegraphics[width=.9\linewidth]{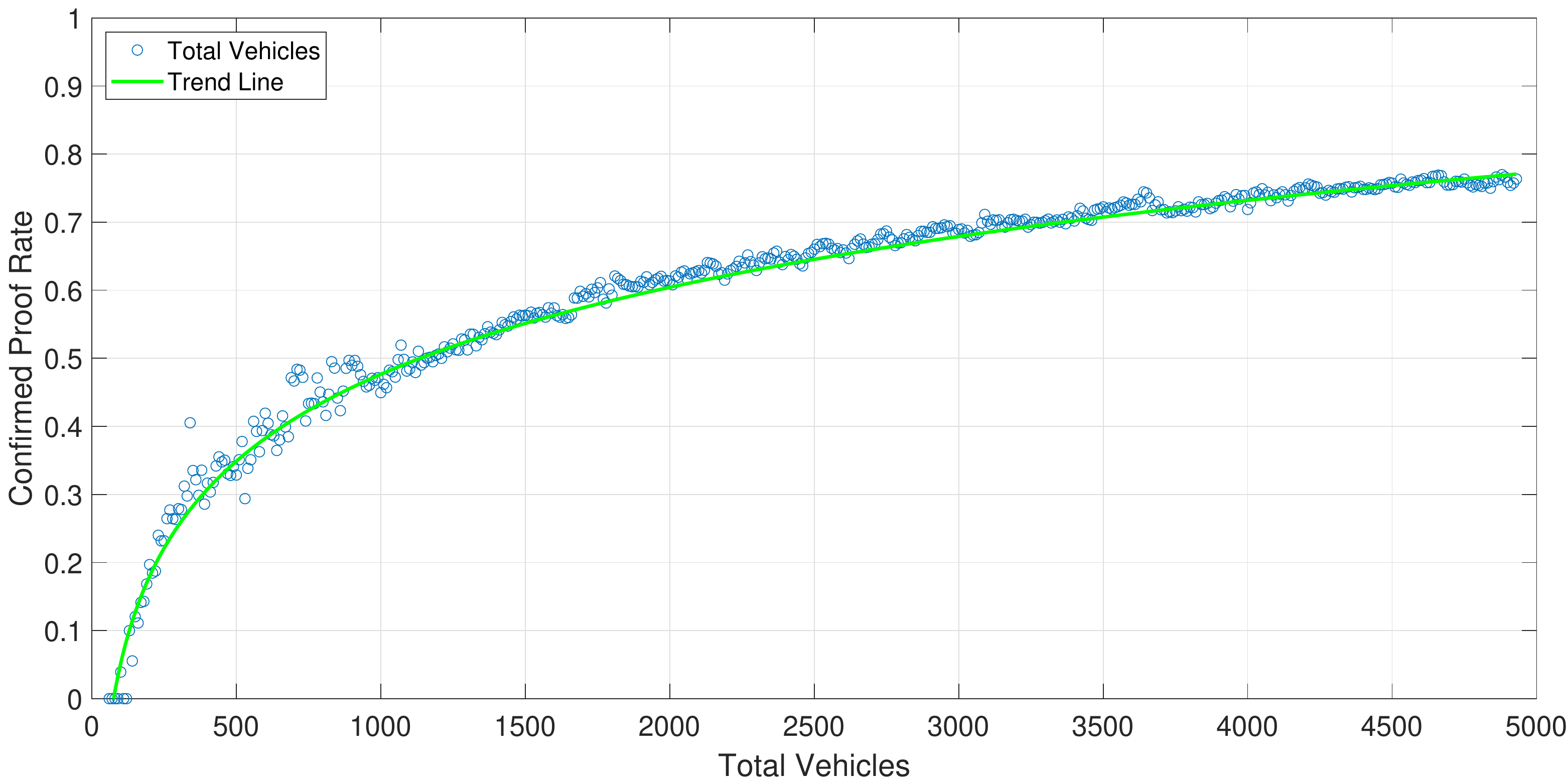}
    \caption{Relation between total vehicles and confirmed proof rate}
    \label{fig:ratio}
\end{figure}

Another interesting discovery is that if we draw the number of total vehicles (including the ones did not observed by any others) as the x-axis and the confirmed proof rate as the y-axis (as shown in figure~\ref{fig:ratio}), the data points emerge a logarithmic-like increase.
Therefore, we use a logarithm-like function (green solid line) to fit the data, and obtained the following results: \begin{equation*}
    y=0.1843\ln{x} - 0.7966
\end{equation*}

The above equation indicates that the confirmed proof rate increases logarithmically with the increase of the total vehicles, which means that the performance of the proposed method should have a decent performance in a dense traffic scenario.
However, since the traffic in Luxembourg is still relatively low, maybe due to its road design, we cannot further increase the simulated traffic while maintaining the realistic traffic model.
Given that logarithm functions do not converge, we suspect this function will not hold in an extreme traffic density scenario.
The reason might be occlusion. In a very dense traffic environment, the distance between two vehicles decreases, which may lead to an increment of occluded area, degrading the overall performance.
Consequent experiments could be conducted in scenarios with high traffic density, e.g., the San Francisco bay area, to inspect this discovery further.

\section{Conclusion}
\label{sec:conclusion}

In this paper, we proposed the zero-knowledge proof of traffic (zk-PoT),
which is a scheme to force the participants to prove the observations to the same vehicles independently,
and convince anyone who obtains more than one proof about the same vehicle.
In the whole procedure, no sensitive data is compromised and the particular vehicle cannot be revealed.
zk-PoT can block most of the common data forging attacks, and collusion attacks can be mitigated.
zk-PoT can be implemented as an extension of the existing cooperative perception standards.
An example is presented using CPS in the ETSI/ITS standard.
zk-PoT can also be integrated with many existing trust models as a strong confidence provider and is especially suitable for the bootstrapping trust establishment process \cite{Hussain2021-ki}.
The key features of zk-PoT are \textit{independent proof}, \textit{cross confirmation}, \textit{public verifiable}, \textit{zero knowledge} and \textit{standard compliant}.

% Beacuse of the scope and length of this study, not all topics could be included.
% In the future, the following issues should be considered:
The following remain as future works:

First, V2V data sharing is not incentivized, which results in vehicles to be selfish to save resources: computation power, communication bandwidth.
This issue is widely discussed in the literature, and the solutions include node-based trust models and social credits.
By analyzing and exploiting the ``privacy-trust dilemma'', an economic model that encourages the vehicles to share their perception in exchange for privacy could be developed.
Based on the model, cross-verified data can produce rewards to the provers, which can be spent to protect their privacy.
Furthermore, if a global economic model is established, the solution to the collusion attack and Sybil attack can be expanded to the whole economic system, which can control their impact globally.

Second, not all vehicles are participating in VANET in the near future, which could result in a degradation of the proposed system because unconnected vehicles do not have their pseudo IDs.
The pseudonym mechanism is crucial in this study because it is the only handy candidate that satisfies two properties, namely ephemeral and length.
As discussed, a non-permanent ID with enough length (i.e., entropy) should be included to protect long-term privacy tracking to the number plate.
To address this problem, we determine other candidates of the entropy source, which may or may not be derived from the non-VANET enabled vehicles themselves.
Besides, by letting the vehicles include the proofs to themselves in the CPM, we can increase the performance of proof matching in sparse traffic situation.
Similar idea can be found in literature \cite{Tsukada2022-ge}.

Finally, some engineering work could be carried out. For example, more attacker models should be implemented in the proof-of-concept to evaluate the performance against the attacks. Also, having an actual implementation based on existing open source ITS projects could be better for illustrating and evaluating the whole system.

\section*{Acknowledgment}
We want to say thank to Dr. Muhammad Asad for giving precious advices on the structure and details of this paper.

These research results were obtained from the commissioned research by the National Institute of Information and Communications Technology (NICT), JAPAN.
\bibliographystyle{unsrt}
\bibliography{main}{}

\end{document}